\documentclass[review,10pt,3p]{elsarticle}

\usepackage{amsmath,amssymb,mathtools}
\usepackage{graphics,subfigure,tabularx}
\usepackage{color}
\usepackage[normalem]{ulem}
\usepackage{stmaryrd}
\usepackage{bm}

\usepackage{booktabs,multirow}
\usepackage[colorlinks,hyperindex,linkcolor=blue,citecolor=blue,urlcolor=blue]{hyperref}
\hypersetup{pdfauthor={R. Huang}}
\newcommand{\figdraft}{false}

\newcommand{\bfe}{\bm{e}}
\newcommand{\bff}{\bm{f}}
\newcommand{\bfF}{\bm{F}}
\newcommand{\bfg}{\bm{g}}
\newcommand{\bfm}{\bm{m}}
\newcommand{\bfM}{\bm{M}}
\newcommand{\bfQ}{\bm{Q}}
\newcommand{\bfR}{\bm{R}}
\newcommand{\bfS}{\bm{S}}
\newcommand{\bfu}{\bm{u}}
\newcommand{\bfx}{\bm{x}}
\newcommand{\bfzero}{\bm{0}}

\newcommand{\dri}{\text{dri}}
\newcommand{\eq}{\text{eq}}
\newcommand{\EOS}{\text{\tiny EOS}}
\newcommand{\intt}{\text{int}}
\newcommand{\noneq}{\text{neq}}

\newcommand{\Err}{E\mspace{-1.5mu}r\mspace{-1.5mu}r}
\newcommand{\pEOS}{p_\EOS^{}}
\newcommand{\psat}{p_\text{sat}^{}}
\newcommand{\rhog}{\rho_g^{}}
\newcommand{\rhol}{\rho_l^{}}

\begin{document}

\begin{frontmatter}
\title{Improved third-order scheme in pseudopotential lattice Boltzmann model for multiphase flows}
\author[CSU]{Rongzong Huang}
\ead{rongzong.huang@csu.edu.cn}
\author[CSU]{Jiayi Huang}
\author[YTZU]{Qing Li}
\address[CSU]{School of Energy Science and Engineering, Central South University, 410083 Changsha, China}
\address[YTZU]{School of Petroleum Engineering, Yangtze University, 430100 Wuhan, China}
\date{April 15, 2026}

\begin{abstract}
    The lattice Boltzmann (LB) equation with a third-order scheme can be regarded as a unified and self-consistent framework of the pseudopotential LB model for multiphase flows. In this work, we theoretically analyze pseudopotential LB simulations of two-phase Poiseuille flow at the discrete level. The finite-difference velocity equation is derived for both grid-aligned and grid-oblique cases. The terms responsible for spurious velocity oscillations near the phase interface are identified. Based on this discrete-level analysis, an improved third-order scheme is proposed to suppress spurious velocity oscillations. This scheme does not introduce any additional conceptual or computational complexity compared with the original one and reduces to the original scheme under static conditions. Numerical simulations of two-phase Poiseuille flow validate the present theoretical analysis and demonstrate the effectiveness of the improved scheme. Then, annular shear flow with a curved phase interface is considered to show that spurious velocity oscillations can also be effectively suppressed by the improved scheme in cases with such interfaces. Finally, the falling of a droplet in a vertical channel is simulated, and the results show that spurious velocity oscillations can lead to an overestimation of the drag force and distinct falling patterns. These results highlight the necessity of using the improved third-order scheme to suppress spurious oscillations and obtain reliable results. 
\end{abstract}


\end{frontmatter}

\section{Introduction} \label{sec.introduction}
Multiphase flows, including those involving phase change, constitute a class of highly important, ubiquitous, and intrinsically complex physical phenomena. They arise in numerous natural processes and play critical roles in a wide range of engineering applications, such as power systems, nuclear engineering, and materials processing. An in-depth understanding of multiphase flow characteristics is of great significance from both scientific and engineering perspectives. Multiphase systems are typically characterized by the dynamic interface evolution, heat and mass transfer across the phase interface, and a wide range of spatial and temporal scales, which render their detailed analysis and accurate prediction extremely challenging. With the rapid advancement of computational hardware and numerical techniques, numerical simulation has become an indispensable tool for studying multiphase flows \cite{Yeoh2019, GarciaVillalba2025}.

As a mesoscopic numerical approach, the lattice Boltzmann (LB) method possesses mesoparticle characteristics and a kinetic-theory background, which enable the natural incorporation of intermolecular interactions \cite{Succi2015, Hosseini2023}. Consequently, it is well-suited for modeling complex multiphase flows from the perspective of underlying physics. Phase interfaces can spontaneously emerge, deform, coalesce, and break up, without requiring additional interface-capturing or tracking techniques. Existing multiphase LB models can be classified into four main categories: the color-gradient model \cite{Gunstensen1991}, the pseudopotential model \cite{Shan1993}, the free-energy model \cite{Swift1995}, and the phase-field model \cite{He1999}. Recently, a multiphase LB model with self-tuning equation of state (EOS) has also been proposed \cite{Huang2019.PRE, Huang2024}, which combines the advantages of the pseudopotential and free-energy models. Among these approaches, the pseudopotential LB model is the simplest one in both concept and implementation and thus has attracted extensive attention, becoming one of the most widely used models for studying multiphase flows \cite{Li2016}.

The pseudopotential LB model was first proposed by Shan and Chen in 1993 \cite{Shan1993}. In this model, a simple pairwise interaction force defined in terms of the pseudopotential is introduced to represent intermolecular interactions. As a result, the non-ideal-gas EOS and non-zero surface tension are simultaneously recovered. The original pseudopotential LB model suffers from several well-known drawbacks, such as a restricted EOS, significant spurious currents, a low achievable density ratio, and the lack of independent adjustment of surface tension. To address these limitations, extensive efforts have been devoted to improving the pseudopotential model. Instead of the original exponential form of the pseudopotential, a square-root form was introduced by He and Doolen \cite{He2002} to incorporate the EOS from thermodynamic theory into the pseudopotential model. Subsequently, Yuan and Schaefer \cite{Yuan2006} systematically examined the numerical performance of various thermodynamic EOSs. Shan \cite{Shan2006} analyzed the mechanism underlying spurious currents near a curved phase interface and proposed highly isotropic gradient operators to suppress them. Wagner and Pooley \cite{Wagner2007} introduced a prefactor into the EOS, which widens the interface thickness in simulations and thus increases the achievable density ratio. Kupershtokh \textit{et al}.\ \cite{Kupershtokh2009} proposed a weighted combination of the pairwise and potential-based interaction forces, enabling the adjustment of the coexistence densities and allowing simulations at lower reduced temperatures. Sbragaglia \textit{et al}.\ \cite{Sbragaglia2007} extended the nearest-neighbor interactions to multirange interactions, which allows the surface tension to be adjusted independently of the EOS.

It is well-known that the pseudopotential LB model inherently suffers from thermodynamic inconsistency. In this model, the coexisting gas and liquid densities ($\rhog$ and $\rhol$) are determined by the mechanical stability condition \cite{Shan2008} 
\begin{equation}
    \int_{\rhog}^{\rhol} (\psat - \pEOS) \dfrac{ \psi^\prime }{ \psi^{1+\epsilon} } d\rho = 0, 
\end{equation}
where $\pEOS$ is the employed EOS, $\psat$ is the saturation pressure, $\psi$ is the pseudopotential with $\psi^\prime = d\psi / d\rho$, and $\epsilon$ is a model-dependent parameter. This condition differs from the Maxwell construction in thermodynamic theory, leading to thermodynamic inconsistency in predicting the coexistence curve. Kupershtokh \textit{et al}.\ \cite{Kupershtokh2009} proposed the exact-difference method (EDM) to incorporate the interaction force into the LB equation. However, both the original Shan-Chen and the EDM forcing schemes yield coexistence curves that depend on the fluid viscosity. In 2012, Li \textit{et al}.\ \cite{Li2012} found that the coexistence curve can be effectively tuned by the parameter $\epsilon$ to approach that predicted by the Maxwell construction, and proposed an improved forcing scheme that provides a degree of freedom to adjust $\epsilon$. Subsequently, a multiple-relaxation-time (MRT) version of this forcing scheme was proposed \cite{Li2013}, enabling simulations at large density ratios. Lycett-Brown and Luo \cite{LycettBrown2015} conducted a third-order truncation error analysis to identify the error terms in the pressure tensor and then proposed an improved forcing scheme. Huang and Wu \cite{Huang2016} carried out a third-order Chapman-Enskog analysis of the MRT pseudopotential LB model, identifying the leading third-order anisotropic and isotropic terms. Based on this analysis, they proposed a self-consistent third-order scheme that allows independent control of both the coexistence curve and the surface tension. Wu \textit{et al}.\ \cite{Wu2020} further extended Huang and Wu’s third-order scheme to three-dimensional pseudopotential LB model.

The aforementioned advances in achieving large density ratios and mitigating thermodynamic inconsistency, together with improvements in numerical stability \cite{Wu2018, Wu2021}, have significantly broadened the applicability of the pseudopotential LB model to complex multiphase flows \cite{Xiong2018, Chen2022, Gong2024, Ezzatneshan2021, Hou2025.JFM}. Recently, Hou \textit{et al}.\ \cite{Hou2025.PRE} examined the performance of various forcing schemes in the pseudopotential LB model under flow conditions and reported spurious velocity oscillations near the phase interface (i.e., the so-called interfacial velocity slip) in simulations of two-phase Poiseuille flow. These oscillations result in significant deviations of the velocity profile. Based on the observation that the standard forcing scheme of Guo \textit{et al}.\ \cite{Guo2002} (corresponding to $\epsilon = 0$) is free of spurious velocity oscillations, and on a conjectured additional term that follows a similar trend as the shear stress across the phase interface, they proposed an improved pseudopotential force to prevent these oscillations. However, this modification sacrifices the clear microscale physical picture of the classical pairwise interaction force. In addition, only the case in which the flow direction and the phase interface are aligned with the lattice grid was examined \cite{Hou2025.PRE}. To identify the origin of spurious velocity oscillations near the phase interface, we theoretically analyze the pseudopotential LB model when simulating two-phase Poiseuille flow at the discrete level and derive the corresponding finite-difference velocity equation. Both the grid-aligned and grid-oblique cases are considered. Based on this discrete-level analysis, we propose an improved third-order scheme to suppress spurious velocity oscillations, in which the classical pairwise interaction force is retained. The remainder of this paper is organized as follows. In Sec.\ \ref{sec.lbm}, a theoretical analysis at the discrete level is conducted, and an improved third-order scheme is proposed. Numerical validations are carried out in Sec.\ \ref{sec.validation}, and a brief conclusion is drawn in Sec.\ \ref{sec.conclusion}.

\section{Pseudopotential LB model} \label{sec.lbm}
The MRT LB equation with a third-order scheme for the density distribution function   can be expressed as \cite{Huang2016}
\begin{subequations}\label{eq.lbe}
    \begin{gather}
        \label{eq.lbe.col}
        \bar{\bfm} (\bfx,t) = \bfm + \delta_t \bfF_m - \bfS \left( \bfm - \bfm^\eq + \dfrac{\delta_t}{2} \bfF_m \right) + \bfS \bfQ_m, \\
        \label{eq.lbe.str}
        f_i (\bfx + \bfe_i \delta_t, t + \delta_t) = \bar{f}_i (\bfx, t),
    \end{gather} 
\end{subequations}
where Eqs.\ (\ref{eq.lbe.col}) and (\ref{eq.lbe.str}) represent the local collision process executed in moment space and the linear streaming process executed in velocity space, respectively, and the overbar denotes the post-collision state. In Eq.\ (\ref{eq.lbe.col}), the coordinates $(\bfx, t)$ are omitted on the right-hand side (RHS) to simplify the notation, $\bfm = \bfM (f_i)^\text{T}$ is the moment with $\bfM$ the transformation matrix, $\bfm^\eq$ is the equilibrium moment, $\bfF_m$ is the discrete force term, $\bfQ_m$ is the discrete source term, and $\bfS$ is the collision matrix. Note that $\bfm^\eq$, $\bfF_m$, and $\bfQ_m$ are of orders $\varepsilon^0$, $\varepsilon^1$, and $\varepsilon^2$, respectively, within a self-consistent Chapman-Enskog analysis of Eq.\ (\ref{eq.lbe}). Here, $\varepsilon$ is the small expansion parameter. In Eq.\ (\ref{eq.lbe.str}), $\bfe_i$ is the discrete velocity, and $\delta_t$ is the time step. The macroscopic density $\rho$ and momentum $\rho \bfu$ are defined as
\begin{equation}\label{eq.rho.rhou}
    \rho = \sum_{i=0}^N {f_i} ,\quad \rho \bfu = \sum_{i=0}^N {\bfe_i f_i} + \dfrac{\delta_t}{2} \bfF,
\end{equation}
where $\bfF$ is the total force composed of the mesoscopic pairwise interaction force and macroscopic external forces (such as gravity). Considering the nearest-neighbor interaction stencil, the pairwise interaction force $\bfF_\intt$ is given as \cite{Shan1993}
\begin{equation}\label{eq.fint}
    \bfF_\intt (\bfx) = -G \psi (\bfx) \sum_{i=1}^N {\omega (|\bfe_i \delta_t|^2) \psi (\bfx + \bfe_i \delta_t) \bfe_i \delta_t },
\end{equation} 
where $G$ is the interaction strength, $\psi$ is the so-called pseudopotential, and $\omega (|\bfe_i \delta_t|^2)$ is the distance-dependent weight to maximize the isotropy degree of $\bfF_\intt$. Through the standard Chapman-Enskog analysis, a non-monotonic EOS can be recovered
\begin{equation}
    \pEOS = \rho c_s^2 + \frac{G\delta _{x}^{2}}{2} \psi^2,
\end{equation}
where $c_s$ is the lattice sound speed. In real applications, a non-ideal-gas EOS (such as the Redlich-Kwong EOS, the Peng-Robinson EOS, the Carnahan-Starling EOS, etc.) can be specified, and then the pseudopotential is inversely calculated via $\psi =\sqrt{ 2 (\pEOS - \rho c_s^2) / G\delta_x^2}$.

To focus on the approach for improving the third-order scheme, the two-dimensional case is considered in this work, while the extension to three dimensions is straightforward. The standard two-dimensional nine-velocity (D2Q9) lattice \cite{Qian1992} and the corresponding orthogonal transformation matrix $\bfM$ \cite{Lallemand2000} are employed. The diagonal MRT collision matrix is 
\begin{equation}
    \bfS =\text{diag} (s_0, s_e, s_\varepsilon, s_j, s_q, s_j, s_q, s_p, s_p) ,
\end{equation}
where $s_p = 1/\tau$ with $\tau$ the dimensionless relaxation time. The equilibrium moment is
\begin{equation}
    \bfm^\eq = \left[ \rho ,\; -2\rho +3\rho |\hat{\bfu}|^2 ,\; \rho -3\rho |\hat{\bfu}|^2 ,\; \rho \hat{u}_x ,\; -\rho \hat{u}_x ,\; \rho \hat{u}_y ,\; -\rho \hat{u}_y ,\; \rho \hat{u}_x^2 - \rho \hat{u}_y^2 ,\; \rho \hat{u}_x \hat{u}_y  \right] ^\text{T},
\end{equation} 
where $\hat{\bfu} = \bfu /c$. The discrete force term is 
\begin{equation}
    \bfF_m = \left[ 0 ,\; 6\hat{\bfF} \cdot \hat{\bfu} ,\; -6\hat{\bfF} \cdot \hat{\bfu} ,\; \hat{F}_x ,\; -\hat{F}_x ,\; \hat{F}_y ,\; -\hat{F}_y ,\; 2\hat{F}_x \hat{u}_x - 2\hat{F}_y \hat{u}_y ,\; \hat{F}_x \hat{u}_y + \hat{F}_y \hat{u}_x \right] ^\text{T},
\end{equation}
where $\hat{\bfF} = \bfF /c$. The discrete source term is expressed as 
\begin{equation}
    \bfQ_m = \left[ Q_{m,0} ,\; Q_{m,1} ,\; Q_{m,2} ,\; Q_{m,3} ,\; Q_{m,4} ,\; Q_{m,5} ,\; Q_{m,6} ,\; Q_{m,7} ,\; Q_{m,8}  \right] ^\text{T},
\end{equation}
where $Q_{m,0} = 0$, $Q_{m,3} = 0$, and $Q_{m,5} = 0$ are required by the fact that the zeroth- and first-order moments $m_0$, $m_3$, and $m_5$ are conserved moments. The elements $Q_{m,1}$, $Q_{m,7}$, and $Q_{m,8}$ for the second-order moments determine the targeted $\varepsilon^3$-order source term \cite{Huang2016}
\begin{equation}
    \bfR_\text{add} = -c^2 \begin{bmatrix}
        \partial_x \big( \tfrac{1}{6} Q_{m,1} + \tfrac{1}{2} Q_{m,7} \big) + \partial_y Q_{m,8} \\
        \partial_x Q_{m,8} + \partial_y \big( \tfrac{1}{6} Q_{m,1} - \tfrac{1}{2} Q_{m,7} \big)\\
    \end{bmatrix} ,
\end{equation} 
which is included in the pressure tensor and used to tune the coexistence curve, or the surface tension, or both. Various kinds of $Q_{m,1}$, $Q_{m,7}$, and $Q_{m,8}$ are proposed explicitly or implicitly \cite{Kupershtokh2009, Li2012, Li2013, Huang2016, Wu2021}. The elements $Q_{m,2}$, $Q_{m,4}$, and $Q_{m,6}$ for the high-order moments are not explicitly involved in the existing theoretical analyses of the pseudopotential LB model. Therefore, their values cannot be unambiguously determined within the existing theoretical framework. In most previous works, $Q_{m,2} = - Q_{m,1}$, $Q_{m,4} = 0$, and $Q_{m,6} = 0$ are artificially imposed, as summarized by Huang \textit{et al}.\ \cite{Huang2019}.

The third-order scheme proposed by Huang and Wu \cite{Huang2016} is adopted as the starting point of the present work because of its unified and self-consistent nature. A standard second-order Chapman-Enskog analysis has been performed for the general case, and a simplified third-order Chapman-Enskog analysis has been performed for the quiescent case. The orders of the variables, derivatives, and terms in these Chapman-Enskog analyses are mutually consistent. Huang and Wu's original discrete source term is \cite{Huang2016}
\begin{equation}
    \bfQ_{m, \text{original}} = \left[ 0 ,\; 3(k_1 + 2k_2) \dfrac{ | \bfF_\intt |^2}{ G \psi^2 c^2 } ,\; -3 (k_1+2k_2) \dfrac{ | \bfF_\intt |^2 }{G \psi^2 c^2 } ,\; 0 ,\; 0 ,\; 0 ,\; 0 ,\; k_1 \dfrac{ F_{\intt,x}^{2} - F_{\intt,y}^{2} }{G \psi^2 c^2 } ,\; k_1 \dfrac{ F_{\intt, x} F_{\intt, y} }{ G \psi^2 c^2 } \right] ^\text{T} ,
\end{equation}
where $k_1$ and $k_2$ are used to independently tune the coexistence curve (i.e., the mechanical stability condition) and surface tension. The parameter $\epsilon$ in the mechanical stability condition is determined by $\epsilon = -8 (k_1 + k_2)$, and the surface tension is tuned by $k_1$. As shown by Hou \textit{et al}.\ \cite{Hou2025.PRE}, this original third-order scheme, as well as most representative forcing schemes \cite{Shan1993, Kupershtokh2009, Li2012}, exhibits spurious velocity oscillations near the phase interface (i.e., the so-called interfacial velocity slip) when simulating two-phase Poiseuille flow. However, the origin of these spurious oscillations remains unclear, and only the grid-aligned case (i.e., the flow direction and the phase interface are parallel to the lattice grid) was considered in Ref.\ \cite{Hou2025.PRE}. In this work, we perform a theoretical analysis of pseudopotential LB simulations of two-phase Poiseuille flow to identify the origin of the spurious oscillations, and then propose an improved discrete source term to suppress them.

\subsection{Theoretical analysis}\label{sec.analysis}
\begin{figure}[tbp]
    \centering
    \includegraphics[scale=1,draft=\figdraft]{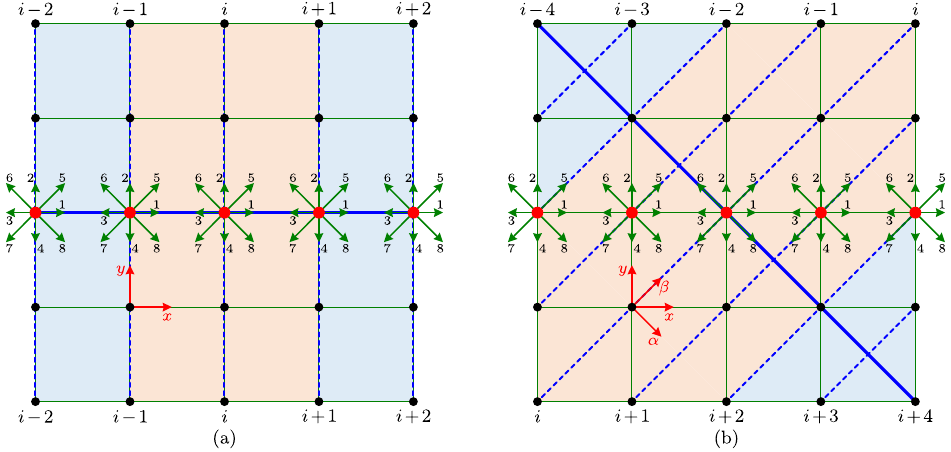}
    \caption[]{Schematics of LB simulations of two-phase Poiseuille flow with the lattice node, lattice grid, and coordinate system shown:\ (a) grid-aligned case, and (b) grid-oblique case. The indices $i$, $i \pm 1$, $i \pm 2$, $\cdots$ represents lattice node positions normal to the flow direction.}
    \label{fig.01}
\end{figure}

The two-phase Poiseuille flow driven by a constant external force $\bfF_\dri$ is analyzed. The driving force $\bfF_\dri$ is applied parallel to the phase interface. Physical quantities, including the density $\rho$, velocity $\bfu$, and pairwise interaction force $\bfF_\intt$, vary only in the direction normal to the flow. Figure \ref{fig.01} illustrates the configurations of pseudopotential LB simulations of this two-phase Poiseuille flow. Both grid-aligned and grid-oblique cases are considered. Without loss of generality, the flow is directed along the $y$ axis for the grid-aligned case, and $45 ^\circ$ clockwise from the $y$ axis (denoted as the $\beta$ axis) for the grid-oblique case. Since Poiseuille flow is a steady problem, the LB equation [i.e., Eq.\ (\ref{eq.lbe})] can be simplified as follows:
\begin{equation}\label{eq.lbe.sim}
    f_i (\bfx + \bfe_i \delta_t) = f_i (\bfx) + \delta_t F_{v,i} (\bfx) - \left[ \bfM^{-1} \bfS \bfM \bff^\noneq (\bfx)- \bfM^{-1} \bfS \bfQ_m (\bfx) \right]_i ,
\end{equation}
where $F_{v,i} = (\bfM^{-1} \bfF_m)_i$ is the discrete force term in velocity space, $\bff^\noneq = \bfM^{-1} \bfm^\noneq$ is the nonequilibrium distribution function with $\bfm^\noneq = \bfm - \bfm^\eq + \tfrac{\delta_t}{2} \bfF_m$ the nonequilibrium moment \cite{Huang2019.JCP}. Based on Huang and Wu's original work, the present improved discrete source term is expressed in the following form: 
\begin{equation}\label{eq.qm.improved}
    \bfQ_{m, \text{improved}} = \left[ 0 ,\; 3(k_1 + 2k_2) \dfrac{ | \bfF_\intt |^2}{ G \psi^2 c^2 } ,\; Q_{m,2} ,\; 0 ,\; Q_{m,4} ,\; 0 ,\; Q_{m,6} ,\; k_1 \dfrac{ F_{\intt,x}^{2} - F_{\intt,y}^{2} }{G \psi^2 c^2 } ,\; k_1 \dfrac{ F_{\intt, x} F_{\intt, y} }{ G \psi^2 c^2 } \right] ^\text{T} .
\end{equation}
The key point is that the elements $Q_{m,4}$ and $Q_{m,6}$ for the third-order moments are not artificially set to zero. On the contrary, they are expected to be of at least order $\varepsilon^2$ and to depend on the velocity $\bfu$. With these constraints, the second- and third-order Chapman-Enskog analyses performed in Ref.\ \cite{Huang2016} are fully applicable to the present improved scheme; therefore, they are not repeated here. In the following, we carry out a discrete-level analysis of LB simulations of two-phase Poiseuille flow. To simplify the notation, the lattice node positions normal to the flow direction are represented by the superscripts $i$, $i \pm 1$, $i \pm 2$, $\cdots$ (see Fig.\ \ref{fig.01}).

\subsubsection{Grid-aligned case}
For the grid-aligned case illustrated in Fig.\ \href{fig.01}{\ref*{fig.01}(a)}, there are 
\begin{equation}\label{eq.ali.1}
    \rho^i = \rho^i (x), \quad
    \begin{dcases}
        \hat{u}_x^i = 0,\\
        \hat{u}_y^i = \hat{u}_y^i (x),\\
    \end{dcases} \quad
    \begin{dcases}
        \hat{F}_{\intt,x}^i = \hat{F}_{\intt,x}^i (x),\\
        \hat{F}_{\intt,y}^i = 0,\\
    \end{dcases} \quad
    \begin{dcases}
        \hat{F}_{\dri,x}^i = 0,\\
        \hat{F}_{\dri,y}^i = \hat{F}_{\dri},\\
    \end{dcases}
\end{equation}
where $\hat{F}_\dri = |\bfF_\dri| /c$  denotes the constant magnitude of the driving force. Based on the definitions of $\rho$ and $\rho \bfu$ [see Eq.\ (\ref{eq.rho.rhou})], the nonequilibrium distribution functions satisfy 
\begin{equation}\label{eq.ali.2}
    \begin{dcases}
        f_{0+1+2+3+4+5+6+7+8}^{\noneq,i} = 0,\\
        f_{1+5+8}^{\noneq,i} - f_{3+6+7}^{\noneq,i} = 0,\\
        f_{2+5+6}^{\noneq,i} - f_{4+7+8}^{\noneq,i} = 0,\\
    \end{dcases}
\end{equation}
where the summation in the subscript represents the summation of the distribution functions over the corresponding directions. For example, $f_{1+5+8}^{\noneq,i} \equiv  f_1^{\noneq,i} + f_5^{\noneq,i} + f_8^{\noneq,i}$.

At lattice node $i$, $f_{2-4}^{i} \equiv f_2^i - f_4^i$ can be determined from the LB equation [i.e., Eq.\ (\ref{eq.lbe.sim})] as follows: 
\begin{equation}\label{eq.ali.3}
    f_{2-4}^{i} = \dfrac{2}{3} \rho^i \hat{u}_y^i + \dfrac{2-s_q}{3 s_q} \delta_t \hat{F}_\dri - \dfrac{1}{3} Q_{m,6}^i,
\end{equation}
where Eqs.\ (\ref{eq.ali.1}) and (\ref{eq.ali.2}) have been used to simplify the derivation. Similarly, from Eq.\ (\ref{eq.lbe.sim}), $f_{5-8}^{i+1} \equiv f_5^{i+1} - f_8^{i+1}$ can be calculated as
\begin{equation}\label{eq.ali.4}
    f_{5-8}^{i+1} = (1-s_p) f_{5-8}^i + \dfrac{s_p}{6} \rho^i \hat{u}_y^i + \dfrac{6 s_q - 4 s_p - s_p s_q}{12 s_q} \delta_t \hat{F}_\dri + \dfrac{2 - s_p}{4} \delta_t \hat{F}_{\intt, x}^i \hat{u}_y^i + \dfrac{s_p}{6} Q_{m,6}^i ,
\end{equation}
and $f_{6-7}^{i-1} \equiv f_6^{i-1} - f_7^{i-1}$ can be calculated as
\begin{equation}\label{eq.ali.5}
    f_{6-7}^{i-1} = (1-s_p) f_{6-7}^i + \dfrac{s_p}{6} \rho^i \hat{u}_y^i + \dfrac{6 s_q - 4 s_p - s_p s_q}{12 s_q} \delta_t \hat{F}_\dri - \dfrac{2 - s_p}{4} \delta_t \hat{F}_{\intt, x}^i \hat{u}_y^i + \dfrac{s_p}{6} Q_{m,6}^i .
\end{equation}

Equation (\ref{eq.ali.3}) is substituted into the definition of $\rho \hat{u}_y$ [see Eq.\ (\ref{eq.rho.rhou})] to obtain 
\begin{equation}\label{eq.ali.6}
    f_{5-8}^i + f_{6-7}^i = \dfrac{1}{3} \rho^i \hat{u}_y^i - \dfrac{4 + s_q}{6 s_q} \delta_t \hat{F}_\dri + \dfrac{1}{3} Q_{m,6}^i .
\end{equation}
By substituting Eq.\ (\ref{eq.ali.6}) into Eq.\ (\ref{eq.ali.4}), shifting the index from $i$ to $i - 1$, and combining the resulting expression with Eq.\ (\ref{eq.ali.5}), we derive 
\begin{equation}\label{eq.ali.7}
    \begin{aligned}
        f_{5-8}^i = & -(1-s_p)^2 f_{6-7}^i - (1-s_p) \left( \dfrac{s_p}{6} \rho^i \hat{u}_y^i - \dfrac{2-s_p}{4} \delta_t \hat{F}_{\intt, x}^i \hat{u}_y^i + \dfrac{s_p}{6} Q_{m,6}^i \right)\\
        & + \left( \dfrac{2-s_p}{6} \rho^{i-1} \hat{u}_y^{i-1} + \dfrac{2-s_p}{4} \delta_t \hat{F}_{\intt, x}^{i-1} \hat{u}_y^{i-1} + \frac{2-s_p}{6} Q_{m,6}^{i-1} \right) + \dfrac{6 s_p s_q - (2 - 2 s_p + s_p^2)(4 + s_q)}{12 s_q} \delta_t \hat{F}_\dri .\\
\end{aligned}
\end{equation}
Similarly, substituting Eq.\ (\ref{eq.ali.6}) into Eq.\ (\ref{eq.ali.5}), shifting the index from $i$ to $i+1$, and combining the resulting expression with Eq.\ (\ref{eq.ali.4}) leads to 
\begin{equation}\label{eq.ali.8}
    \begin{aligned}
        f_{6-7}^i = & -(1-s_p)^2 f_{5-8}^i - (1-s_p) \left( \dfrac{s_p}{6} \rho^i \hat{u}_y^i + \dfrac{2-s_p}{4} \delta_t \hat{F}_{\intt,x}^i \hat{u}_y^i + \dfrac{s_p}{6} Q_{m,6}^i \right)\\
        & + \left( \dfrac{2-s_p}{6} \rho^{i+1} \hat{u}_y^{i+1} - \dfrac{2-s_p}{4} \delta_t \hat{F}_{\intt, x}^{i+1} \hat{u}_y^{i+1} + \dfrac{2-s_p}{6} Q_{m,6}^{i+1} \right) +\dfrac{6 s_p s_q - (2 - 2 s_p + s_p^2) (4 + s_q)}{12 s_q} \delta_t \hat{F}_\dri .\\
    \end{aligned}
\end{equation}
Adding Eqs.\ (\ref{eq.ali.7}) and (\ref{eq.ali.8}), and then substituting Eq.\ (\ref{eq.ali.6}), we finally obtain the following finite-difference velocity equation: 
\begin{equation}\label{eq.fde.ali}
    \left( \rho^{i-1} + \dfrac{3}{2} \delta_t \hat{F}_{\intt, x}^{i-1} \right) \hat{u}_y^{i-1} + \left( \rho^{i+1} - \dfrac{3}{2} \delta_t \hat{F}_{\intt, x}^{i+1} \right) \hat{u}_y^{i+1} - 2 \rho^i \hat{u}_y^i + \left( Q_{m,6}^{i-1} + Q_{m,6}^{i+1} - 2Q_{m,6}^i \right) = -\dfrac{ \hat{F}_\dri }{\nu} \delta_x^2 ,
\end{equation}
where $\nu = c_s^2 \delta_t \left( s_p^{-1} - 0.5 \right)$ is the kinematic viscosity. For single-phase flows, $\rho = \text{const}$, $\bfF_\intt = \bfzero$, and $\bfQ_m = \bfzero$. Equation (\ref{eq.fde.ali}) then reduces to the second-order finite-difference form of the Navier-Stokes equations. The corresponding solution yields an exact parabolic profile as long as the nonslip condition is satisfied.

For two-phase Poiseuille flow, both $\rho$ and $\hat{F}_{\intt, x}$ vary with the lattice node index $i$. To explicitly identify the origin of spurious velocity oscillations and to derive an appropriate choice of $Q_{m,6}$, we further simplify Eq.\ (\ref{eq.fde.ali}). For this purpose, summing Eq.\ (\ref{eq.lbe.sim}) over the directions $0$, $2$, and $4$ yields 
\begin{equation}\label{eq.ali.9}
    f_{0+2+4}^i = \dfrac{2}{3} \rho^i - \dfrac{\epsilon}{8} \dfrac{\delta_t^2 \hat{F}_{\intt, x}^{i,2} }{ \hat{\psi}^{i,2} } ,
\end{equation}
where $\hat{\psi} = \sqrt{-G \delta_t^2} \psi$ is the rescaled pseudopotential, and $s_e = s_p$ is set to simplify the algebraic manipulations. Such a choice is acceptable, considering that the spurious oscillations also exist for the single-relaxation-time (SRT) collision matrix with $s_e = s_p$, and that the improved $\bfQ_m$ should be independent of the relaxation parameters. By substituting Eq.\ (\ref{eq.ali.9}) into the definitions of $\rho$ and $\rho \bfu$ [see Eq.\ (\ref{eq.rho.rhou})], we can derive 
\begin{subequations}\label{eq.ali.10}
    \begin{gather}
        f_{1+5+8}^i = \dfrac{1}{6} \rho^i - \dfrac{1}{4} \delta_t \hat{F}_{\intt,x}^i + \dfrac{\epsilon}{16} \dfrac{\delta_t^2 \hat{F}_{\intt, x}^{i,2} }{ \hat{\psi}^{i,2} }, \\
        f_{3+6+7}^i = \dfrac{1}{6} \rho^i + \dfrac{1}{4} \delta_t \hat{F}_{\intt, x}^i + \dfrac{\epsilon}{16} \dfrac{\delta_t^2 \hat{F}_{\intt, x}^{i,2} }{ \hat{\psi}^{i,2} }.
    \end{gather}
\end{subequations}
Meanwhile, summing Eq.\ (\ref{eq.lbe.sim}) separately over the directions $1$, $5$, and $8$, and over the directions $3$, $6$, and $7$, yields 
\begin{subequations}\label{eq.ali.11}
    \begin{gather}
        f_{1+5+8}^{i+1} = f_{1+5+8}^i + \dfrac{1}{2} \delta_t \hat{F}_{\intt, x}^i , \\
        f_{3+6+7}^{i-1} = f_{3+6+7}^i - \dfrac{1}{2} \delta_t \hat{F}_{\intt, x}^i .
    \end{gather}
\end{subequations}
By shifting the index in Eq.\ (\ref{eq.ali.10}) and substituting the resulting expressions, together with Eq.\ (\ref{eq.ali.10}), into Eq.\ (\ref{eq.ali.11}), we can derive 
\begin{subequations}\label{eq.ali.12}
    \begin{gather}
        \rho^{i+1} - \dfrac{3}{2} \delta_t \hat{F}_{\intt,x}^{i+1} = \rho^i + \dfrac{3}{2} \delta_t \hat{F}_{\intt,x}^i + \dfrac{3 \epsilon}{8} \dfrac{\delta_t^2 \hat{F}_{\intt,x}^{i,2} }{ \hat{\psi}^{i,2} } - \dfrac{3 \epsilon}{8} \dfrac{ \delta_t^2 \hat{F}_{\intt, x}^{i+1,2} }{ \hat{\psi}^{i+1,2} }, \\
        \rho^{i-1} + \dfrac{3}{2} \delta_t \hat{F}_{\intt,x}^{i-1} = \rho^i - \dfrac{3}{2} \delta_t \hat{F}_{\intt,x}^i + \dfrac{3 \epsilon}{8} \dfrac{\delta_t^2 \hat{F}_{\intt,x}^{i,2} }{ \hat{\psi}^{i,2} } - \dfrac{3 \epsilon}{8} \dfrac{ \delta_t^2 \hat{F}_{\intt, x}^{i-1,2} }{ \hat{\psi}^{i-1,2} } .
    \end{gather}
\end{subequations}
With Eq.\ (\ref{eq.ali.12}), the finite-difference velocity equation [i.e., Eq.\ (\ref{eq.fde.ali})] can be further reformulated as 
\begin{equation}\label{eq.ali.final}
    \begin{aligned}
        & \left( \rho^i + \dfrac{3 \epsilon}{8} \dfrac{\delta_t^2 \hat{F}_{\intt, x}^{i,2} }{ \hat{\psi}^{i,2} } \right) \left( \hat{u}_y^{i-1} + \hat{u}_y^{i+1} - 2\hat{u}_y^{i} \right) + \dfrac{3}{2} \delta_t \hat{F}_{\intt, x}^i \left( \hat{u}_y^{i+1} - \hat{u}_y^{i-1} \right) \\
        &\mspace{50mu} -\dfrac{3 \epsilon}{8} \left( \dfrac{ \delta_t^2 \hat{F}_{\intt, x}^{i-1,2} }{ \hat{\psi}^{i-1,2} } \hat{u}_y^{i-1} + \dfrac{ \delta_t^2 \hat{F}_{\intt, x}^{i+1,2} }{ \hat{\psi}^{i+1,2} } \hat{u}_y^{i+1} - 2 \dfrac{ \delta_t^2 \hat{F}_{\intt, x}^{i,2} }{ \hat{\psi}^{i,2} } \hat{u}_y^i \right) + \left( Q_{m,6}^{i-1} + Q_{m,6}^{i+1} - 2 Q_{m,6}^i \right) = -\dfrac{ \hat{F}_\dri }{\nu} \delta_x^2 .
    \end{aligned}
\end{equation}
It can be observed from Eq.\ (\ref{eq.ali.final}) that spurious velocity oscillations near the phase interface are induced by the third term on the left-hand side (LHS), since the spurious oscillations vanish when $\epsilon = 0$ (as shown by Hou \textit{et al}.\ \cite{Hou2025.PRE}), while the prefactor $\rho^i + 3 \epsilon \delta_t^2 \hat{F}_{\intt, x}^{i,2}  \big/ (8 \hat{\psi}^{i,2})$ in the first term can be interpreted as an equivalent density. This conclusion can be readily confirmed by directly solving Eq.\ (\ref{eq.ali.final}) with a given two-phase density profile. To eliminate the spurious oscillations, we can set 
\begin{equation}\label{eq.qm6.ali}
    Q_{m,6} = \dfrac{3\epsilon}{8} \dfrac{ \delta_t^2 \hat{F}_{\intt, x}^2 }{ \hat{\psi}^2 } \hat{u}_y .
\end{equation}
Note that Eq.\ (\ref{eq.ali.final}) is derived for the grid-aligned case, and thus the above $Q_{m,6}$ is strictly applicable only to the grid-aligned case. For the grid-oblique case, the derived finite-difference velocity equation may be different, and the corresponding $Q_{m,4}$ and/or $Q_{m,6}$ would also change.

\subsubsection{Grid-oblique case}
For the grid-oblique case illustrated in Fig.\ \href{fig.01}{\ref*{fig.01}(b)}, the velocities in $\alpha$ and $\beta$ directions are calculated as 
\begin{subequations}\label{eq.obl.1}
    \begin{gather}
        \rho \hat{u}_\alpha \equiv \dfrac{\sqrt{2}}{2} (\rho \hat{u}_x - \rho \hat{u}_y ) = \dfrac{ \sqrt{2} }{2} (f_{1+4} - f_{2+3} + 2 f_{8-6} ) + \dfrac{\delta_t}{2} \hat{F}_{\alpha} , \\
        \label{eq.obl.1b}
        \rho \hat{u}_\beta \equiv \dfrac{\sqrt{2}}{2} (\rho \hat{u}_x + \rho \hat{u}_y ) = \dfrac{\sqrt{2}}{2} (f_{1-4} + f_{2-3} + 2f_{5-7} ) + \dfrac{\delta _t}{2} \hat{F}_{\beta} ,
    \end{gather}
\end{subequations}
where $\hat{F}_{\alpha} \equiv \sqrt{2} (\hat{F}_x - \hat{F}_y) \big/ 2$ and $\hat{F}_{\beta} \equiv \sqrt{2} (\hat{F}_x + \hat{F}_y) \big/ 2$ are the total forces in $\alpha$ and $\beta$ directions, respectively. Due to geometric symmetry, there are $\hat{u}_x^i = \hat{u}_y^i$, $\hat{F}_{\intt, x}^i = -\hat{F}_{\intt, y}^i$, and $\hat{F}_{\dri, x}^i = \hat{F}_{\dri, y}^i$, and thus the following relations hold 
\begin{equation}\label{eq.obl.2}
    \rho^i = \rho^i (\alpha), \quad
    \begin{dcases}
        \hat{u}_{\alpha}^i = 0, \\
        \hat{u}_{\beta}^i = \hat{u}_{\beta}^i (\alpha),\\
    \end{dcases} \quad
    \begin{dcases}
        \hat{F}_{\intt, \alpha}^i = \hat{F}_{\intt, \alpha}^i (\alpha), \\
        \hat{F}_{\intt, \beta}^i = 0, \\
    \end{dcases} \quad
    \begin{dcases}
        \hat{F}_{\dri, \alpha}^i = 0, \\
        \hat{F}_{\dri, \beta}^i = \hat{F}_\dri. \\
    \end{dcases}
\end{equation}
Based on the definitions of $\rho$ and $\rho \bfu$ [see Eqs.\ (\ref{eq.rho.rhou}) and (\ref{eq.obl.1})], the nonequilibrium distribution functions satisfy 
\begin{equation}\label{eq.obl.3}
    \begin{dcases}
        f_{0+1+2+3+4+5+6+7+8}^{\noneq, i} = 0, \\
        f_{1+4}^{\noneq, i} - f_{2+3}^{\noneq, i} + 2 f_{8-6}^{\noneq, i} = 0, \\
        f_{1-4}^{\noneq, i} + f_{2-3}^{\noneq, i} + 2 f_{5-7}^{\noneq, i} = 0. \\
    \end{dcases}
\end{equation}

At lattice node $i$, $f_{5-7}^i \equiv f_5^i - f_7^i$ can be determined from the LB equation [i.e., Eq.\ (\ref{eq.lbe.sim})] as follows:
\begin{equation}\label{eq.obl.4}
    f_{5-7}^i = \sqrt{2} \dfrac{1}{6} \rho^i \hat{u}_\beta^i + \sqrt{2} \dfrac{2-s_q}{12 s_q} \delta_t \hat{F}_\dri + \dfrac{1}{6} Q_{m,4+6}^i ,
\end{equation} 
where $Q_{m,4+6}^i \equiv Q_{m,4}^i + Q_{m,6}^i$, and Eqs.\ (\ref{eq.obl.2}) and (\ref{eq.obl.3}) have been used for simplification. Similarly, from Eq.\ (\ref{eq.lbe.sim}), $f_{1-4}^{i+1} \equiv f_1^{i+1} - f_4^{i+1}$ can be calculated as 
\begin{equation}\label{eq.obl.5}
    f_{1-4}^{i+1} = (1-s_p) f_{1-4}^i + \sqrt{2} \dfrac{s_p}{3} \rho^i \hat{u}_\beta^i + \sqrt{2} \dfrac{3 s_q - s_p - s_p s_q}{6 s_q} \delta_t \hat{F}_\dri + \dfrac{2-s_p}{2} \delta_t \hat{F}_{\intt, \alpha}^i \hat{u}_\beta^i - \dfrac{s_p}{6} Q_{m,4+6}^i, 
\end{equation}
and $f_{2-3}^{i-1} \equiv f_2^{i-1}-f_3^{i-1}$ can be calculated as 
\begin{equation}\label{eq.obl.6}
    f_{2-3}^{i-1} = (1-s_p) f_{2-3}^i + \sqrt{2} \dfrac{s_p}{3} \rho^i \hat{u}_\beta^i + \sqrt{2} \dfrac{3 s_q - s_p - s_p s_q}{6 s_q} \delta_t \hat{F}_\dri - \dfrac{2 - s_p}{2} \delta_t \hat{F}_{\intt, \alpha}^i \hat{u}_\beta^i - \dfrac{s_p}{6} Q_{m,4+6}^i .
\end{equation}

Equation (\ref{eq.obl.4}) is substituted into the definition of $\rho \hat{u}_\beta$ [see Eq.\ (\ref{eq.obl.1b})] to obtain 
\begin{equation}\label{eq.obl.7}
    f_{1-4}^i + f_{2-3}^i = \sqrt{2} \dfrac{2}{3} \rho^i \hat{u}_\beta^i - \sqrt{2} \dfrac{1 + s_q}{3 s_q} \delta_t \hat{F}_\dri - \dfrac{1}{3} Q_{m,4+6}^i .
\end{equation}
By substituting Eq.\ (\ref{eq.obl.7}) into Eq.\ (\ref{eq.obl.5}), shifting the index from $i$ to $i - 1$, and combining the resulting expression with Eq.\ (\ref{eq.obl.6}), we derive 
\begin{equation}\label{eq.obl.8}
    \begin{aligned}
        f_{1-4}^i = & - (1-s_p)^2 f_{2-3}^i - (1-s_p) \left( \sqrt{2} \dfrac{s_p}{3} \rho^i \hat{u}_\beta^i - \dfrac{2 - s_p}{2} \delta_t \hat{F}_{\intt, \alpha}^i \hat{u}_\beta^i - \dfrac{s_p}{6} Q_{m,4+6}^i \right) \\
        & + \left( \sqrt{2} \dfrac{2 - s_p}{3} \rho^{i-1} \hat{u}_\beta^{i-1} + \frac{2-s_p}{2} \delta_t \hat{F}_{\intt, \alpha}^{i-1} \hat{u}_\beta^{i-1} - \dfrac{2 - s_p}{6} Q_{m,4+6}^{i-1} \right) \\
        & + \sqrt{2} \dfrac{3 s_p s_q - (2 - 2 s_p + s_p^2)(1 + s_q)}{6 s_q} \delta_t \hat{F}_\dri .\\
    \end{aligned}
\end{equation}
Similarly, substituting Eq.\ (\ref{eq.obl.7}) into Eq.\ (\ref{eq.obl.6}), shifting the index from $i$ to $i+1$, and combining the resulting expression with Eq.\ (\ref{eq.obl.5}) leads to 
\begin{equation}\label{eq.obl.9}
    \begin{aligned}
        f_{2-3}^i = & - (1-s_p)^2 f_{1-4}^i - (1-s_p) \left( \sqrt{2} \dfrac{s_p}{3} \rho^i \hat{u}_\beta^i + \dfrac{2 - s_p}{2} \delta_t \hat{F}_{\intt, \alpha}^i \hat{u}_\beta^i - \dfrac{s_p}{6} Q_{m,4+6}^i \right) \\
        & + \left( \sqrt{2} \dfrac{2 - s_p}{3} \rho^{i+1} \hat{u}_\beta^{i+1} - \dfrac{2 - s_p}{2} \delta_t \hat{F}_{\intt, \alpha}^{i+1} \hat{u}_\beta^{i+1} - \dfrac{2 - s_p}{6} Q_{m,4+6}^{i+1} \right) \\
        & + \sqrt{2} \dfrac{3 s_p s_q - (2 - 2 s_p + s_p^2) (1 + s_q)}{6 s_q} \delta_t \hat{F}_\dri .\\
    \end{aligned}
\end{equation}
By adding Eqs.\ (\ref{eq.obl.8}) and (\ref{eq.obl.9}), substituting Eq.\ (\ref{eq.obl.7}), and using $\hat{F}_{\intt, \alpha} = \sqrt{2} \hat{F}_{\intt, x}$, we finally obtain the following finite-difference velocity equation: 
\begin{equation}\label{eq.fde.obl}
    \left( \rho^{i-1} + \dfrac{3}{2} \delta_t \hat{F}_{\intt, x}^{i-1} \right) \hat{u}_\beta^{i-1} + \left( \rho^{i+1} - \dfrac{3}{2} \delta_t \hat{F}_{\intt, x}^{i+1} \right) \hat{u}_\beta^{i+1} - 2 \rho^i \hat{u}_\beta^i - \dfrac{\sqrt{2}}{4} \left( Q_{m,4+6}^{i-1} + Q_{m,4+6}^{i+1} - 2 Q_{m,4+6}^i \right) = - \dfrac{\hat{F}_\dri}{\nu} \dfrac{\delta_x^2}{2} ,
\end{equation}
which corresponds to Eq.\ (\ref{eq.fde.ali}) for the grid-aligned case. However, it is difficult to further simplify Eq.\ (\ref{eq.fde.obl}), as in the grid-aligned case, to derive the required $Q_{m,4+6}$, because the streaming processes at lattice node $i$ in the directions normal to the flow originate from lattice nodes $i \pm 2$, rather than $i \pm 1$, as illustrated in Fig.\ \href{fig.01}{\ref*{fig.01}(b)}. Fortunately, Eq.\ (\ref{eq.fde.obl}) has the same form as Eq.\ (\ref{eq.fde.ali}); therefore, the required $Q_{m,4+6}^{i}$ for the grid-oblique case can be heuristically derived from that for the grid-aligned case [see Eq.\ (\ref{eq.qm6.ali})].

The factors multiplying $Q_{m,4+6}$ in Eq.\ (\ref{eq.fde.obl}) and $Q_{m,6}$ in Eq.\ (\ref{eq.fde.ali}) differ, even after accounting for $\sqrt{2} /2$ arising from the transformation between the $x\text{-}y$ and $\alpha \text{-} \beta$ coordinate systems. Note that $\delta_x^2 \big/ 2$ on the RHS of Eq.\ (\ref{eq.fde.obl}) arises from the distance between two adjacent lattice nodes in the direction normal to the flow being $\delta_x \big/ \sqrt{2}$. For the nearest-neighbor interaction given by Eq.\ (\ref{eq.fint}), $\hat{F}_{\intt, x}^i$ in Eqs.\ (\ref{eq.fde.ali}) and (\ref{eq.fde.obl}) can be simplified to 
\begin{equation}\label{eq.obl.10}
    \hat{F}_{\intt, x}^i =
    \begin{dcases}
        \hat{\psi}^i \dfrac{ \hat{\psi}^{i+1} - \hat{\psi}^{i-1} }{2} \dfrac{1}{\delta_t}, & \text{grid-aligned} ,\\
        \hat{\psi}^i \dfrac{ \hat{\psi}^{i+2} + 4 \hat{\psi}^{i+1} - 4 \hat{\psi}^{i-1} - \hat{\psi}^{i-2} }{12} \dfrac{1}{\delta_t}, & \text{grid-oblique} , 
    \end{dcases}
\end{equation}
implying that Eq.\ (\ref{eq.fde.obl}) employs a seven-point finite-difference stencil, whereas Eq.\ (\ref{eq.fde.ali}) uses a five-point stencil. Owing to the above differences between Eqs.\ (\ref{eq.fde.obl}) and (\ref{eq.fde.ali}), the $Q_{m,6}$ for the grid-aligned case [see Eq.\ (\ref{eq.qm6.ali})] cannot be directly extended to $Q_{m, 4+6}$ for the grid-oblique case. Nevertheless, inspired by Eq.\ (\ref{eq.qm6.ali}), $Q_{m, 4+6}$ should take the following form: 
\begin{equation}\label{eq.obl.11}
    Q_{m,4+6} = - \dfrac{4}{\sqrt{2}} K \dfrac{ \delta_t^2 \hat{F}_{\intt, x}^{2}}{ \hat{\psi}^2 } \hat{u}_\beta = -2K \dfrac{ \delta_t^2 \hat{F}_{\intt, x}^2}{ \hat{\psi}^2 } (\hat{u}_x + \hat{u}_y) ,
\end{equation}
where the factor $K$ is determined to be $K = (24 \epsilon - 15) /32$ by directly solving Eq.\ (\ref{eq.fde.obl}) with a given two-phase density profile to suppress spurious velocity oscillations. Based on Eq.\ (\ref{eq.obl.11}), and considering the symmetry relation between the third-order moments $m_4$ and $m_6$, as well as the geometric symmetry, the corresponding $Q_{m,4}$ and $Q_{m,6}$ are chosen as 
\begin{equation}\label{eq.qm46.obl}
    Q_{m,4} = \dfrac{15 - 24\epsilon}{16} \dfrac{ \delta_t^2 \hat{F}_{\intt, x}^{2} }{ \hat{\psi}^2 } \hat{u}_x, \quad
    Q_{m,6} = \dfrac{15 - 24\epsilon}{16} \dfrac{ \delta_t^2 \hat{F}_{\intt, y}^{2} }{ \hat{\psi}^2 } \hat{u}_y.
\end{equation}
As expected, the above $Q_{m,6}$ for the grid-oblique case differs from that for the grid-aligned case [see Eq.\ (\ref{eq.qm6.ali})].

\subsection{Summary}
To ensure compatibility with Eq.\ (\ref{eq.qm6.ali}) for the grid-aligned case and Eq.\ (\ref{eq.qm46.obl}) for the grid-oblique case, and by substituting $\hat{\psi} = \sqrt{-G \delta_t^2} \psi$, $\hat{\bfF}_\intt = \bfF_\intt /c$, and $\hat{\bfu} = \bfu /c$, the improved $Q_{m,4}$ and $Q_{m,6}$ can be expressed in the following general form: 
\begin{equation}\label{eq.qm46.improved}
    Q_{m,4} = \left( \dfrac{30 \epsilon - 15}{16} \dfrac{ F_{\intt,x}^2 }{ G \psi^2 c^2 } - \dfrac{3 \epsilon}{8} \dfrac{ F_{\intt,y}^2 }{ G \psi^2 c^2 } \right) \dfrac{u_x}{c} , \quad 
    Q_{m,6} = \left( \dfrac{30 \epsilon - 15}{16} \dfrac{ F_{\intt,y}^2 }{ G \psi^2 c^2 } - \dfrac{3 \epsilon}{8} \dfrac{ F_{\intt,x}^2 }{ G \psi^2 c^2 } \right) \dfrac{u_y}{c} .
\end{equation}
In the Chapman-Enskog analysis, the pairwise interaction force $\bfF_\intt$ is of order $\varepsilon^1$, whereas the macroscopic velocity $\bfu$ and the pseudopotential $\psi$ are of order $\varepsilon^0$. Consequently, the above improved $Q_{m,4}$ and $Q_{m,6}$ are of order $\varepsilon^2$. Furthermore, under static conditions (i.e., $\bfu = \bfzero$), the above improved $Q_{m,4}$ and $Q_{m,6}$ vanish. Thus, the Chapman-Enskog analyses performed in Ref.\ \cite{Huang2016}, including the standard second-order analysis for the general case and the simplified third-order analysis for the quiescent case, are fully applicable to the present improved scheme. Here, it is worth noting that $Q_{m,2}$ for the fourth-order moment remains undetermined after the present analysis aimed at eliminating spurious velocity oscillations. Based on the forms of the improved $Q_{m,4}$ and $Q_{m,6}$, we heuristically choose $Q_{m,2} = - Q_{m,1} \big/2$ in the present improved scheme. Moreover, compared with the original scheme, the present improved scheme does not introduce any additional conceptual or computational complexity.

\section{Numerical validations} \label{sec.validation}
Numerical tests are conducted in this section to validate the theoretical analysis and the improved third-order scheme. In the simulations, the canonical van der Waals (vdW) EOS is adopted as a representative example, since the maximum density ratio in LB simulations depends on the interface thickness rather than the EOS \cite{Li2023}. The vdW EOS is given by 
\begin{equation}
    \pEOS = K_\EOS \left( \dfrac{\rho RT}{1 - b\rho} - a \rho^2 \right) ,
\end{equation}
where the vdW constants $a = 9/49$ and $b = 2/21$, and the gas constant $R = 1$. The scaling factor $K_\EOS$ is used to adjust the interface thickness \cite{Wagner2007} and is set to $K_\EOS = 1/16$ in the present simulations. The critical temperature and density are $T_\text{cr} = 8a/(27Rb)$ and $\rho _\text{cr} = 1/(3b)$. The reduced temperature is defined as $T_r = T / T_\text{cr}$. The nonslip condition is treated by the improved nonequilibrium extrapolation scheme \cite{Huang2019.JCP}, and the wettability condition is handled by the virtual-density scheme \cite{Benzi2006}. The lattice spacing and time step are set as $\delta_x = 1$ and $\delta_t = 1$. The relaxation parameters in $\bfS$ are chosen according to $s_0 = s_j = 1$, $s_e = s_p$, $s_\varepsilon = s_e$, and $\big( s_q^{-1} - 1/2 \big) \big( s_p^{-1} - 1/2 \big) = 1/12$. Without loss of generality, the parameters $k_{1,2}$ in $\bfQ_m$ are set to $k_1 = k_2 = -\epsilon / 16$.

\subsection{Plane Poiseuille flow}
\begin{figure}[tbp]
    \centering
    \includegraphics[scale=1,draft=\figdraft]{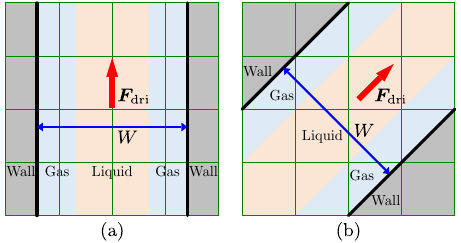}
    \caption[]{Schematics of two-phase Poiseuille flow driven by a constant force $\bfF_\dri$ between two parallel planes of width $W$:\ (a) grid-aligned case, and (b) grid-oblique case.}
    \label{fig.02}
\end{figure}

First, the plane Poiseuille flow, theoretically analyzed in Sec.\ \ref{sec.analysis}, is simulated on a $256 \delta_x \times 256 \delta_x$ domain. Figure \ref{fig.02} shows the geometric configurations for the grid-aligned and grid-oblique cases. The distance between the two parallel planes is fixed at $W = 128 \sqrt{2} \delta_x$. Gas layers with a thickness of $W/4$ are placed near each walls, while a liquid layer with a thickness of $W/2$ occupies the middle region. The driving force is fixed at $|\bfF_\dri| = 2.0 \times 10^{-7}$. The parameter $\epsilon$ in the mechanical stability condition is chosen as $0$, $1$, and $2$, respectively, and the corresponding reduced temperatures $T_r$ are set to $0.72$, $0.65$, and $0.54$. These choices ensure that the liquid-to-gas density ratio in the simulations remains close to $60$. Two dimensionless relaxation times, $\tau = 0.8$ and $1.5$, are employed to examine the $\tau$-dependence of the numerical results.

\begin{figure}[tbp]
    \centering
    \includegraphics[scale=1,draft=\figdraft]{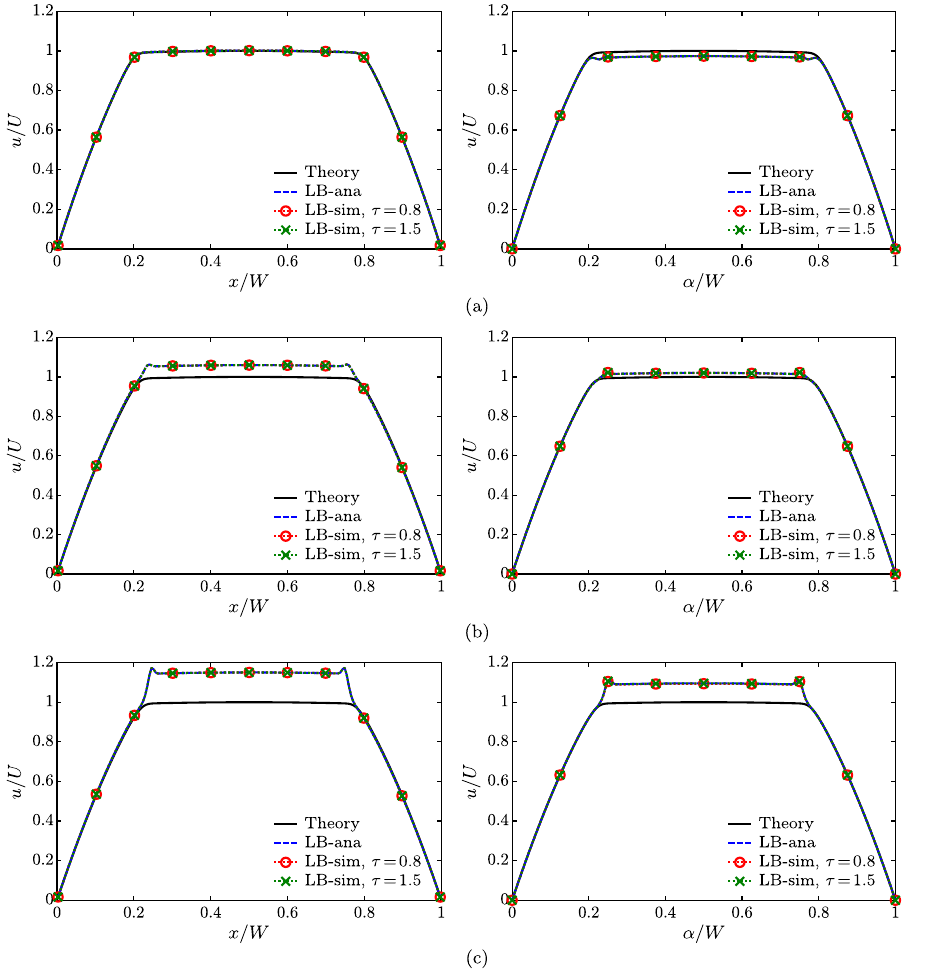}
    \caption[]{Velocity profiles of two-phase Poiseuille flow obtained from the LB equation with the original third-order scheme for (a) $\epsilon = 0$, (b) $\epsilon = 1$, and (c) $\epsilon = 2$. The left and right panels correspond to the grid-aligned and grid-oblique cases, respectively.}
    \label{fig.03}
\end{figure}

Figure \ref{fig.03} shows the velocity profiles obtained from the LB equation with the original third-order scheme, where $u$ denotes the velocity component parallel to the wall, and $U$ is the maximum value of the theoretical result. For the plane Poiseuille flow, the macroscopic Navier-Stokes equations can be simplified to 
\begin{equation}\label{eq.u.plane}
    \dfrac{\partial}{\partial x} \left( \rho \nu \dfrac{\partial u}{\partial x} \right) = - | \bfF_\dri |.
\end{equation}
The theoretical result (denoted by ``Theory'') is obtained by integrating Eq.\ (\ref{eq.u.plane}) using the converged density profile from the LB simulation. Similarly, the analytical result of the LB equation (denoted by ``LB-ana'') is obtained by solving the derived finite-difference velocity equation [i.e., Eq.\ (\ref{eq.fde.ali}) or (\ref{eq.fde.obl})] with the same converged density profile. As shown in Fig.\ \ref{fig.03}, the analytical results of the LB equation are in good agreement with the LB simulation results (denoted by ``LB-sim''), thereby validating the derivations presented in Sec.\ \ref{sec.analysis}. For the grid-aligned case (see the left panel), the velocity profile obtained with the original third-order scheme coincides with the theoretical profile when $\epsilon = 0$ (i.e., $\bfQ_{m, \text{original}} = \bfzero$), but exhibits spurious oscillations near the phase interface when $\epsilon = 1$ and $2$ (i.e., $\bfQ_{m, \text{original}} \neq \bfzero$). As $\epsilon$ increases, the spurious velocity oscillations become more pronounced. For the grid-oblique case (see the right panel), the velocity profile obtained with the original third-order scheme always exhibits spurious oscillations and deviates markedly from the theoretical profile within the liquid layer, even when $\bfQ_{m, \text{original}} = \bfzero$. Such a finding suggests that the analysis of Hou \textit{et al}.\ \cite{Hou2025.PRE} and their improved scheme may not remain valid when the phase interface is not aligned with the lattice grid. Figure \ref{fig.03} also shows that the spurious velocity oscillations are independent of the dimensionless relaxation time $\tau$.

\begin{figure}[tbp]
    \centering
    \includegraphics[scale=1,draft=\figdraft]{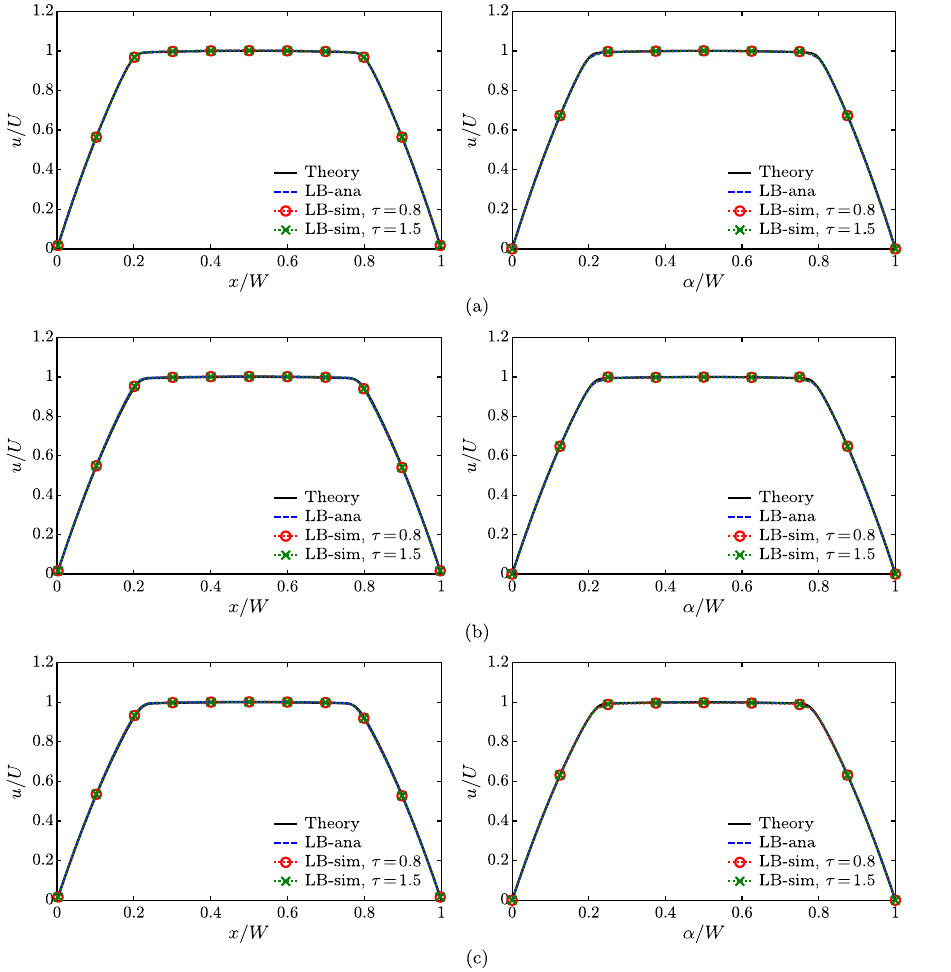}
    \caption[]{Velocity profiles of two-phase Poiseuille flow obtained from the LB equation with the improved third-order scheme for (a) $\epsilon = 0$, (b) $\epsilon = 1$, and (c) $\epsilon = 2$. The left and right panels correspond to the grid-aligned and grid-oblique cases, respectively.}
    \label{fig.04}
\end{figure}

Figure \ref{fig.04} shows the velocity profiles obtained from the LB equation with the improved third-order scheme [see Eqs.\ (\ref{eq.qm.improved}) and (\ref{eq.qm46.improved})]. The analytical results of the LB equation are in good agreement with the LB simulation results, further confirming the derivations presented in Sec.\ \ref{sec.analysis}. For the grid-aligned case (see the left panel), the velocity profile obtained with the improved third-order scheme varies smoothly across the phase interface and coincides with the theoretical profile for $\epsilon = 0$, $1$, and $2$. For the grid-oblique case (see the right panel), it exhibits only subtle spurious oscillations near the phase interface and remains in good agreement with the theoretical profile, even within the liquid layer. The results in Fig.\ \ref{fig.04} confirm the identified origin of the spurious velocity oscillations (see Sec.\ \ref{sec.analysis}) and demonstrate their effective suppression by the improved third-order scheme. Moreover, Fig.\ \ref{fig.04} shows that this suppression is independent of the dimensionless relaxation time $\tau$.

For quantitative comparisons between the original and improved third-order schemes, the relative error between the LB simulation result and the theoretical result is calculated and defined as 
\begin{equation}
    \Err = \sqrt{ \dfrac{ \sum{ (u_\text{LB-sim}^{} - u_\text{theory}^{})^2 } }{ \sum{ u_\text{theory}^2 } }}.
\end{equation}
Table \ref{table.1} lists the relative errors $\Err$ for the cases shown in Figs.\ \ref{fig.03} and \ref{fig.04}. As seen in Table \ref{table.1}, the improved third-order scheme yields much smaller $\Err$ due to the effective suppression of the spurious velocity oscillations. As $\epsilon$ increases from $0$ to $2$, $\Err$ remains at essentially the same level for both the grid-aligned and grid-oblique cases. In contrast, for the original third-order scheme, $\Err$ increases markedly once the spurious velocity oscillations appear (vanishing only for the grid-aligned case with $\epsilon = 0$, and present in all other cases) and generally grows with increasing $\epsilon$. Table \ref{table.1} confirms that the relative error $\Err$ obtained with both the original and improved third-order schemes is nearly independent of the dimensionless relaxation time $\tau$ for the grid-aligned case, while exhibiting only a very weak dependence on $\tau$ for the grid-oblique case.

\begin{table}[tbp]    
    \centering
    \caption{Comparison of the relative errors $\Err$ in LB simulations of two-phase Poiseuille flow between the original and improved third-order schemes.}\label{table.1}
    \label{Table.1}
    \setlength{\tabcolsep}{12.9pt} %
    \begin{tabular}{cccccc}
        \toprule
        \multicolumn{2}{c}{\multirow{2}{*}{Cases}}    & \multicolumn{2}{c}{Original}                    & \multicolumn{2}{c}{Improved} \\
                                                        \cmidrule(lr){3-4}                                \cmidrule(lr){5-6}
                                       &              & $\tau = 0.8$           & $\tau = 1.5$           & $\tau = 0.8$           & $\tau = 1.5$ \\
        \midrule
        \multirow{2}{*}{$\epsilon = 0$} & Grid-aligned & $0.123 \times 10^{-2}$ & $0.123 \times 10^{-2}$ & $0.123 \times 10^{-2}$ & $0.123 \times 10^{-2}$ \\
                                        & Grid-oblique & $2.361 \times 10^{-2}$ & $2.373 \times 10^{-2}$ & $0.182 \times 10^{-2}$ & $0.189 \times 10^{-2}$ \vspace{1.5ex} \\
        \multirow{2}{*}{$\epsilon = 1$} & Grid-aligned & $5.221 \times 10^{-2}$ & $5.222 \times 10^{-2}$ & $0.415 \times 10^{-2}$ & $0.416 \times 10^{-2}$ \\
                                        & Grid-oblique & $1.734 \times 10^{-2}$ & $1.779 \times 10^{-2}$ & $0.252 \times 10^{-2}$ & $0.233 \times 10^{-2}$ \vspace{1.5ex} \\
        \multirow{2}{*}{$\epsilon = 2$} & Grid-aligned & $12.922\times 10^{-2}$ & $12.923\times 10^{-2}$ & $0.389 \times 10^{-2}$ & $0.389 \times 10^{-2}$ \\
                                        & Grid-oblique & $8.025 \times 10^{-2}$ & $8.185 \times 10^{-2}$ & $0.398 \times 10^{-2}$ & $0.293 \times 10^{-2}$ \\
        \bottomrule
    \end{tabular}
\end{table}

\begin{figure}[tbp]
    \centering
    \includegraphics[scale=1,draft=\figdraft]{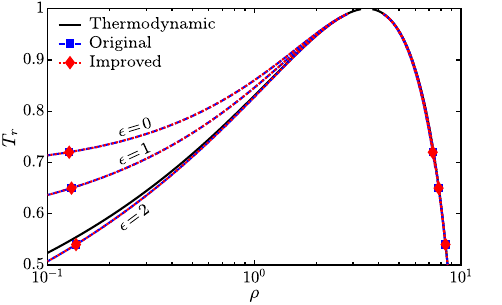}
    \caption[]{Comparison of coexistence curves obtained from LB simulations of two-phase Poiseuille flow in the grid-oblique case with a dimensionless relaxation time $\tau = 1.5$ between the original and improved third-order schemes. The symbols correspond to the cases shown in Figs.\ \ref{fig.03} and \ref{fig.04}.}
    \label{fig.05}
\end{figure}

Before proceeding further, the influence of the improved third-order scheme on the coexistence curve is examined. The grid-oblique case is considered with the dimensionless relaxation time fixed at $\tau = 1.5$. The liquid-to-gas density ratio is varied by adjusting the reduced temperature. Meanwhile, the magnitude of the driving force is tuned to maintain a low-Mach-number condition. The coexisting densities are then determined from the converged density profile. Figure \ref{fig.05} shows the coexistence curves obtained from the LB equation with the original and improved third-order schemes for $\epsilon = 0$, $1$, and $2$. The thermodynamic curve determined from the Maxwell construction is also included for comparison. As shown in Fig.\ \ref{fig.05}, the coexistence curves obtained with the original and improved third-order schemes are indistinguishable. Figure \ref{fig.05} also confirms that the coexistence curve can be effectively tuned by the parameter $\epsilon$ in the mechanical stability condition, and that a value of $\epsilon$ close to $2$ should be adopted in practice to achieve the thermodynamic consistency in predicting the coexistence curve \cite{Li2012}.

\subsection{Annular shear flow}
\begin{figure}[tbp]
    \centering
    \includegraphics[scale=1,draft=\figdraft]{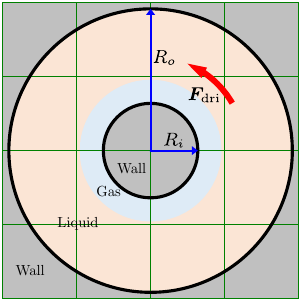}
    \caption[]{Schematic of two-phase annular shear flow driven by an angular force $\bfF_\dri$ between two concentric circles with inner and outer radii $R_i$ and $R_o$.}
    \label{fig.06}
\end{figure}

The two-phase annular shear flow is then considered (see Fig.\ \ref{fig.06}) to examine the effectiveness of the improved third-order scheme in cases with curved phase interfaces. This annular shear flow is driven by an angular force within two concentric circles with inner and outer radii $R_i = 64 \delta_x$ and $R_o = 256 \delta_x$, respectively. Unlike the two-phase plane Poiseuille flow shown in Fig.\ \ref{fig.02}, only a gas layer of thickness $(R_o - R_i)/4$ is placed near the inner wall here, while the liquid occupies the remaining region to avoid the Rayleigh-Taylor instability in the centrifugal field. Furthermore, the magnitude of the driving force is prescribed to vary with the radial coordinate as $|\bfF_\dri| = F_o r^2 / R_o^2$ with $F_o = 3.0 \times 10^{-5}$, thereby preventing excessively large velocity in the gas layer. The parameter $\epsilon$ in the mechanical stability condition is set to $1.895$ to achieve thermodynamic consistency in predicting the coexistence curve. The reduced temperature is set to $T_r = 0.5$, corresponding to a liquid-to-gas density ratio of approximately $113$. The dimensionless relaxation time is fixed at $\tau = 0.8$.

\begin{figure}[tbp]
    \centering
    \includegraphics[scale=1,draft=\figdraft]{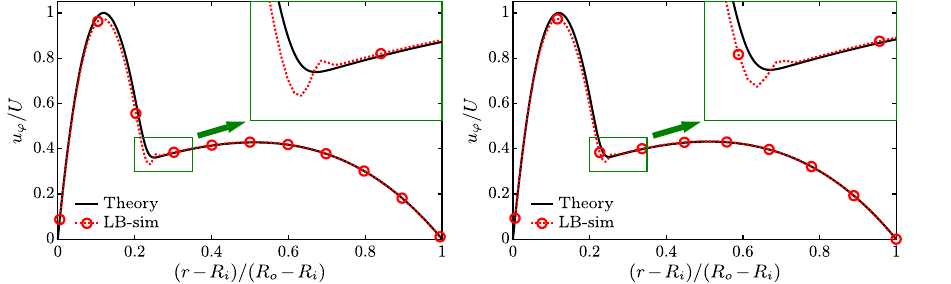}
    \caption[]{Velocity profiles of two-phase annular shear flow obtained from the LB equation with the original third-order scheme. The left and right panels correspond to the results in the horizontal and $45^\circ$-inclined directions, respectively.}
    \label{fig.07}
\end{figure}

\begin{figure}[tbp]
    \centering
    \includegraphics[scale=1,draft=\figdraft]{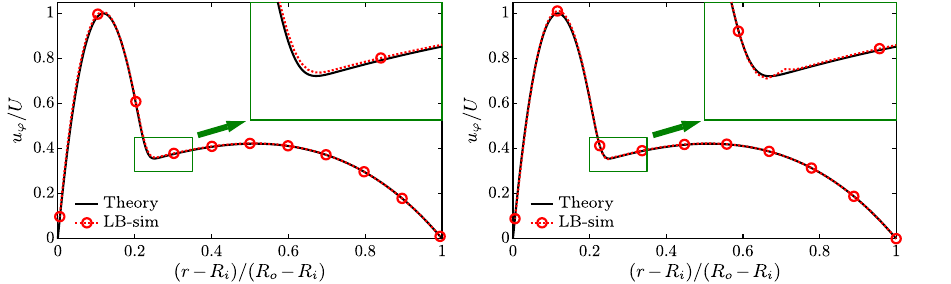}
    \caption[]{Velocity profiles of two-phase annular shear flow obtained from the LB equation with the improved third-order scheme. The left and right panels correspond to the results in the horizontal and $45^\circ$-inclined directions, respectively.}
    \label{fig.08}
\end{figure}

Figures \ref{fig.07} and \ref{fig.08} show the velocity profiles obtained from the LB equation with the original and improved third-order schemes, respectively. Here, $u_\varphi$ denotes the tangential velocity, and $U$ is its maximum value in the theoretical solution. Results in both the horizontal and $45^\circ$-inclined directions are included. For this two-phase annular shear flow, the Navier-Stokes equations can be simplified to 
\begin{equation}\label{eq.u.annular}
    \dfrac{1}{r^2} \dfrac{\partial}{\partial r} \left[ \rho \nu r^3 \dfrac{\partial (r^{-1} u_\varphi)}{\partial r} \right] = -|\bfF_\dri| .
\end{equation}
Using the converged density profile from the LB simulation, the theoretical velocity profile is obtained by integrating Eq.\ (\ref{eq.u.annular}). Here, it is worth noting that the radial density distributions in the horizontal and $45^\circ$-inclined directions differ slightly in the LB simulation; accordingly, the corresponding theoretical velocity profiles should be calculated separately. As shown in Fig.\ \ref{fig.07}, the result obtained from the LB equation with the original third-order scheme exhibits pronounced spurious velocity oscillations near the phase interface in both the horizontal and $45^\circ$-inclined directions, leading to a significant velocity deviation within the gas layer. In contrast, the result obtained with the improved third-order scheme (see Fig.\ \ref{fig.08}) varies smoothly across the phase interface in the horizontal direction and shows only subtle oscillations in the $45^\circ$-inclined directions (see the inserted enlarged view). Despite these weak oscillations, the velocity profile remains in good agreement with the theoretical result in both the gas and liquid layers. These comparisons demonstrate that the improved third-order scheme, derived from the theoretical analysis of plane Poiseuille flow with a planar phase interface, remains effective in suppressing spurious velocity oscillations in cases with curved phase interfaces.

\subsection{Droplet falling in vertical channel}
\begin{figure}[tbp]
    \centering
    \includegraphics[scale=1,draft=\figdraft]{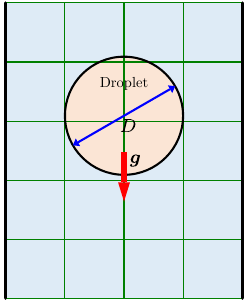}
    \caption[]{Schematic of a droplet falling in an infinite vertical channel under gravitational acceleration $\bfg$.}
    \label{fig.09}
\end{figure}

To demonstrate the influence of spurious velocity oscillations near the phase interface in practical applications, the falling of a liquid droplet in an infinitely long vertical channel under gravity is considered, as illustrated in Fig.\ \ref{fig.09}. The channel width is set to $W = 256 \delta_x$, and a liquid droplet with a diameter $D = W/2$ is initially placed at the center of the channel. To mimic an infinitely long vertical channel, the height of the computational domain is set to a sufficiently large value of $H = 2048 \delta_x$, and periodic boundary conditions are imposed in the $y$ direction. The virtual wall density is fixed at $\rho_s^{} = \rhog$ to prevent droplet sticking to the wall. The parameter $\epsilon$ in the mechanical stability condition is set to $1.895$, and the reduced temperature is $T_r = 0.5$. The dimensionless relaxation time is fixed at $\tau = 0.8$. The simulation is first run without gravity until a converged state is reached, providing a suitable initial condition for the falling process. The gravitational body force $\bfF_\text{gra} = (\rho -\rhog) \bfg$, with $|\bfg| = 0.5 \times 10^{-7}$, $1.0 \times 10^{-7}$, and $2.0 \times 10^{-7}$, respectively, is then imposed, and the droplet begins to fall.

\begin{figure}[tbp]
    \centering
    \includegraphics[scale=1,draft=\figdraft]{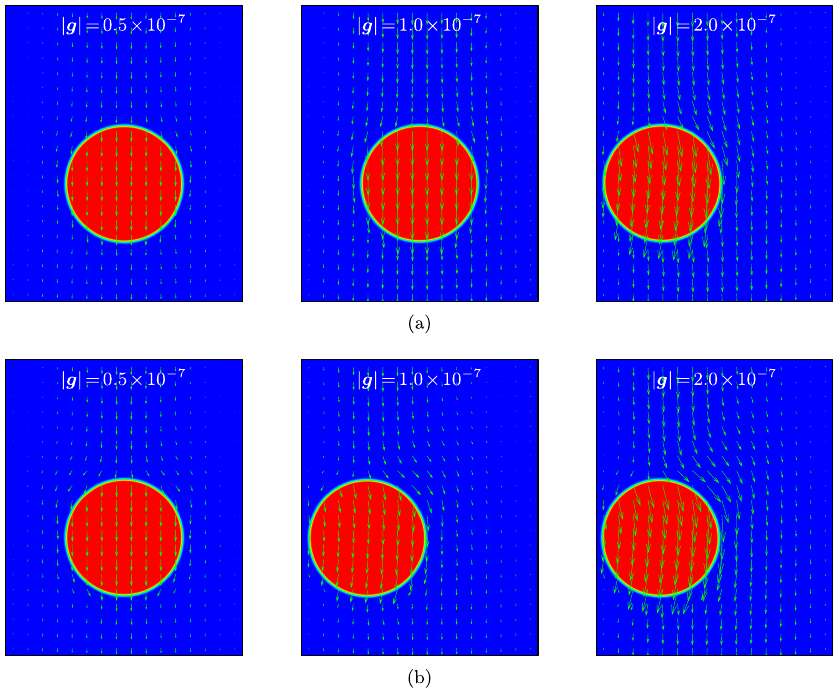}
    \caption[]{Snapshots of the falling droplet and the corresponding velocity field in the quasi-steady state obtained from the LB equation with (a) the original third-order scheme and (b) the improved third-order scheme. The gravitational acceleration is $|\bfg| = 0.5 \times 10^{-7}$, $1.0 \times 10^{-7}$, and $2.0 \times 10^{-7}$ from the left, middle, and right panels, respectively.}
    \label{fig.10}
\end{figure}

Figure \ref{fig.10} shows snapshots of the falling droplet and the corresponding velocity field in the quasi-steady state obtained from the LB equation with the original and improved third-order schemes, respectively. For comparison, the droplet is shifted to the same vertical position in each snapshot. As shown in Fig.\ \ref{fig.10}, when $|\bfg| = 0.5 \times 10^{-7}$, the droplet falls along the centerline of the channel for both the original and improved third-order schemes. The velocity vectors inside the droplet are uniform in both direction and magnitude, indicating that the droplet undergoes a purely translational motion. When the gravitational acceleration increases to $|\bfg| = 1.0 \times 10^{-7}$, the falling pattern obtained with the original third-order scheme remains essentially the same as that at $|\bfg| = 0.5 \times 10^{-7}$, except for a higher falling velocity of the droplet (see below for the definition). In contrast, for the improved third-order scheme, the droplet moves close to the channel wall, and the velocity vectors inside the droplet are no longer uniform in either direction or magnitude, indicating a combined translational and rotational motion. Moreover, the falling velocity of the droplet is smaller than that obtained with the original third-order scheme. As the gravitational acceleration further increases to $|\bfg| = 2.0 \times 10^{-7}$, the falling patterns obtained with both schemes become essentially identical to that obtained with the improved scheme at $|\bfg| = 1.0 \times 10^{-7}$. Moreover, the falling velocity of the droplet shows no visible difference between the two schemes, although the droplet remains slightly closer to the channel wall with the improved scheme. Figure \ref{fig.10} demonstrates that spurious velocity oscillations near the phase interface in the original third-order scheme can lead to different two-phase flow patterns in practical applications, highlighting the necessity of the present improved third-order scheme.

\begin{figure}[tbp]
    \centering
    \includegraphics[scale=1,draft=\figdraft]{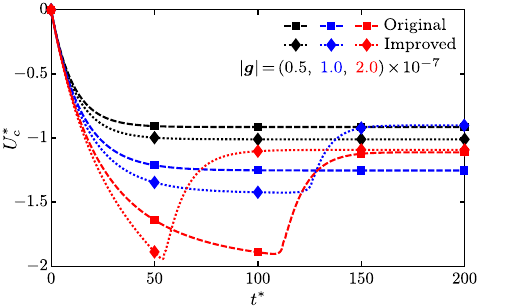}
    \caption[]{Temporal variations of the vertical velocity of the mass center of the falling droplet, obtained from the LB equation with the original and improved third-order schemes. Black, blue, and red correspond to $|\bfg| = 0.5 \times 10^{-7}$, $1.0 \times 10^{-7}$, and $2.0 \times 10^{-7}$, respectively.}
    \label{fig.11}
\end{figure}

For further comparison between the original and improved third-order schemes, the falling velocity of the droplet (i.e., the vertical velocity of the droplet center of mass) is calculated as 
\begin{equation}
    U_c = \dfrac{ {\displaystyle \int} \rho u_y dA }{ {\displaystyle \int} \rho dA },
\end{equation}
where the integration is performed over the liquid region. The dimensionless falling velocity $U_c^\ast$ and time $t^\ast$ are defined as 
\begin{equation}
    U_c^\ast = \dfrac{ U_c }{ \sqrt{ |\bfg| D (\rhol - \rhog) \big/ \rhog } }, \quad
    t^\ast = \dfrac{t}{ \sqrt{ D \big/ |\bfg| } } .
\end{equation}
Figure \ref{fig.11} shows the variation of $U_c^\ast$ with $t^\ast$ obtained from the LB equation with the original and improved third-order schemes. As the gravitational acceleration increases, the droplet exhibits a stronger acceleration during the initial stage ($t^\ast \leq 50$). During this stage, the droplet acceleration obtained with the improved scheme is larger than that with the original scheme. When $|\bfg| = 0.5 \times 10^{-7}$, the droplet acceleration gradually decreases with increasing $t^\ast$, and the quasi-steady state is eventually reached. The terminal falling velocity obtained with the improved third-order scheme is about $10\%$ larger than that with the original scheme ($|U_c^\ast| = 1.0104$ versus $0.9144$). When the gravitational acceleration increases to $|\bfg| = 1.0 \times 10^{-7}$, the variation of $U_c^\ast$ with $t^\ast$ is essentially the same as that at $|\bfg| = 0.5 \times 10^{-7}$ for both the original and improved schemes when $t^\ast \leq 120$. After that, the droplet continues to fall along the centerline of the channel for the original scheme [see the middle panel of Fig.\ \href{fig.10}{\ref*{fig.10}(a)}]. In contrast, for the improved scheme, the droplet rapidly migrates toward the channel wall [see the middle panel of Fig.\ \href{fig.10}{\ref*{fig.10}(b)}] once the centerline falling becomes unstable at sufficiently large falling velocities. As the droplet approaches the wall, the falling velocity significantly decreases due to the increased drag induced by wall effects. Consequently, the terminal falling velocity obtained with the improved third-order scheme is about $28\%$ smaller than that with the original scheme ($|U_c^\ast| = 0.9015$ versus $1.2535$). When $|\bfg| = 2.0 \times 10^{-7}$, the droplet first falls along the channel centerline and then migrates toward the channel wall for both schemes. The migration is triggered at approximately half the time for the improved scheme compared with the original scheme ($t^\ast \approx 53.8$ versus $107.6$), while the falling velocities at the onset of migration are close to each other ($|U_c^\ast| \approx 1.92$). The falling velocity subsequently decreases as the droplet approaches the channel wall. It is also noted that the terminal falling velocity with the improved scheme is slightly smaller than that with the original scheme ($|U_c^\ast| = 1.0928$ versus $1.1116$). The results in Fig.\ \ref{fig.11} indicate that spurious velocity oscillations near the phase interface in the original third-order scheme lead to an overestimation of the drag force in the droplet falling process. Therefore, the present improved third-order scheme is essential for suppressing these oscillations and obtaining reliable results.

\section{Conclusion} \label{sec.conclusion}
In this work, the pseudopotential LB model is theoretically analyzed at the discrete level in the context of two-phase Poiseuille flow simulations. The finite-difference velocity equation is derived for both grid-aligned and grid-oblique cases. The terms responsible for spurious velocity oscillations near the phase interface are clearly identified. These oscillations are shown to arise when $\epsilon \neq 0$ in the grid-aligned case and are always present in the grid-oblique case. Based on this discrete-level analysis, an improved third-order scheme is proposed to suppress spurious velocity oscillations. In this scheme, the source terms for the third-order moments are determined by the velocity and the squared components of the pairwise interaction force, rather than being artificially set to zero. As a result, the conceptual and computational advantages of the pseudopotential LB model are fully preserved. Numerical simulations of two-phase Poiseuille flow are first performed to validate both the theoretical analysis at the discrete level and the effectiveness of the improved third-order scheme. Then, annular shear flow is simulated to demonstrate the effectiveness of the improved scheme in cases with curved phase interfaces. Finally, the falling of a liquid droplet in an infinitely long vertical channel is simulated to illustrate the influence of spurious velocity oscillations in practice. The results indicate that such oscillations lead to an overestimation of the drag force and result in distinct falling patterns. These findings highlight the necessity of the improved third-order scheme for suppressing spurious velocity oscillations and obtaining reliable simulation results.

\section*{Acknowledgements}
This work was supported by the National Natural Science Foundation of China through Grant No.\ 52376086.

\biboptions{sort&compress}
\bibliographystyle{elsarticle-num}

\end{document}